\documentclass[a4paper,english,runningheads]{llncs}
\usepackage[T1]{fontenc}
\usepackage[latin9]{inputenc}
\usepackage{xcolor}
\usepackage{units}
\usepackage{amsmath}
\usepackage{amssymb}
\usepackage[all]{xy}
\PassOptionsToPackage{normalem}{ulem}
\usepackage{ulem}

\makeatletter

\providecommand{\tabularnewline}{\\}
\providecolor{lyxadded}{rgb}{0,0,1}
\providecolor{lyxdeleted}{rgb}{1,0,0}

\usepackage[all]{xy}

\newcommand{\LFT}{\mathsf{LFT}}
\newcommand{\JDC}{\mathsf{JDC}}
\newcommand{\SI}{\mathsf{SI}}

\newcommand{\D}{\mathrm{D}}

\makeatother

\usepackage{babel}
\begin{document}

\title{Quantum Entanglement and the Issue of Selective Influences in Psychology:
An Overview}

\titlerunning{Selective Influences and Quantum Entanglement}

\author{Ehtibar N. Dzhafarov\textsuperscript{1}\and Janne V. Kujala\textsuperscript{2}}

\authorrunning{E. N. Dzhafarov and J. V. Kujala}

\institute{\textsuperscript{1}Purdue University\\
ehtibar@purdue.edu\\
\textsuperscript{2}University of Jyv\"askyl\"a\\
 jvk@iki.fi\\
}

\titlerunning{Quantum Entanglement and Selective Influences }

\tocauthor{E.N. Dzhafarov and J.V. Kujala}
\maketitle
\begin{abstract}
Similar formalisms have been independently developed in psychology,
to deal with the issue of selective influences (deciding which of
several experimental manipulations selectively influences each of
several, generally non-independent, response variables), and in quantum
mechanics (QM), to deal with the EPR entanglement phenomena (deciding
whether an EPR experiment allows for a ``classical'' account). The
parallels between these problems are established by observing that
any two noncommuting measurements in QM are mutually exclusive and
can therefore be treated as analogs of different values of one and
the same input. Both problems reduce to that of the existence of a
jointly distributed system of random variables, one variable for every
value of every input (in psychology) or every measurement on every
particle involved (in an EPR experiment). We overview three classes
of necessary conditions (some of them also sufficient under additional
constraints) for the existence of such joint distributions. 

\end{abstract}

\section{Introduction}

Given a set of inputs into a system and a set of stochastically non-independent
outputs, what is the precise meaning and means of ascertaining that
a given output \emph{is not influenced} by a given input? This paper
reviews the developments related to this question. 

The problem can be illustrated on the following \emph{diagram of selective
influences}: 

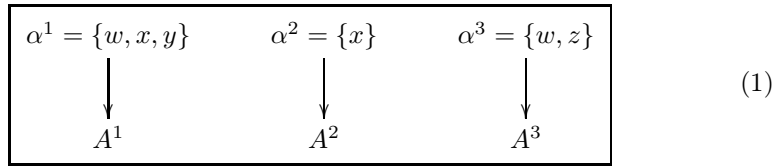
\emph{
\begin{equation}
\begin{array}{c}
\boxed{\xymatrix{\alpha^{1}=\left\{ w,x,y\right\} \ar[d] & \alpha^{2}=\left\{ x\right\} \ar[d] & \alpha^{3}=\left\{ w,z\right\} \ar[d]\\
A^{1} & A^{2} & A^{3}
}
}\end{array}
\end{equation}
}$A^{1}$\emph{, $A^{2}$,} and $A^{3}$ here are \emph{random outputs},
$w,x,y,z$ are \emph{inputs} (usually referred to as \emph{external
factors} in psychology and as \emph{measurement settings} in QM),
and arrows indicate the relation ``may influence'': thus, the diagram
does not say that $A^{2}$ is necessarily influenced by $x$, but
rather that $A^{2}$ is not influenced by $w,y,z$. The diagram is
shown in the \emph{canonical form}, i.e., the inputs are redefined,
$\left\{ w,x,y\right\} $ into $\alpha^{1}$, $\left\{ x\right\} $
into $\alpha^{2}$, etc., so that each output $A^{i}$ may only be
influenced by a single input $\alpha^{i}$ that may not influence
other outputs. We say then, for brevity, that $\left(A^{1},A^{2},A^{3}\right)$
are \emph{selectively influenced} by $\left(\alpha^{1},\alpha^{2},\alpha^{3}\right)$
and write this as 
\begin{equation}
\left(A^{1},A^{2},A^{3}\right)\looparrowleft\left(\alpha^{1},\alpha^{2},\alpha^{3}\right).
\end{equation}
Inputs $\left(\alpha^{1},\alpha^{2},\alpha^{3}\right)$ are treated
as deterministic quantities, i.e., even if they are random variables,
the joint distribution of the outputs is always conditioned on their
specific values. Each input can have one of several values, and the
joint distribution of $\left(A^{1},A^{2},A^{3}\right)$ is known for
each \emph{allowable treatment}, a\emph{ }combination of input values.
Thus, if $w,x,y,z$ are all binary, then $\alpha^{1},\alpha^{2},\alpha^{3}$
may be viewed as inputs with 8, 2, and 4 values, respectively, but
the number of allowable treatments cannot exceed $16<8\times2\times4$.
It can be less than 16 because some of the combinations may be physically
impossible or simply not used or observed. 

As a motivating example, consider a double-detection experiment in
which two stimuli, say brief flashes, are presented simultaneously
(right-left) or in a succession (first-second), each on one of two
levels of intensity. The observer is asked to state, for each of the
two \emph{observation areas} (i.e., locations or time intervals),
whether it contains a flash (Yes/No). The results of such an experiment
are statistical estimates of 16 probabilities
\begin{equation}
p\left(A^{1},A^{2}|\alpha^{1},\alpha^{2}\right)=\Pr\left[A^{1}:\left\{ \begin{array}{c}
Yes\\
No
\end{array}\right.,A^{2}:\left\{ \begin{array}{c}
Yes\\
No
\end{array}\right.\left|\;\alpha^{1}:\left\{ \begin{array}{c}
\alpha_{1}^{1}\\
\alpha_{2}^{1}
\end{array}\right.,\alpha^{2}:\left\{ \begin{array}{c}
\alpha_{1}^{2}\\
\alpha_{2}^{2}
\end{array}\right.\right.\right],\label{eq:2-2-2}
\end{equation}
where $\alpha^{i}$ ($i=1,2)$ is the input representing the $i$th
observation area (with values $\alpha_{1}^{i},\alpha_{2}^{i}$), and
$A^{i}$ is the response (Yes or No) to the $i$th observation area.
Assume that $A^{1}$ and $A^{2}$ for a given $\left(\alpha_{i}^{1},\alpha_{j}^{2}\right)$
are not independent (due to attention fluctuations, perceptual learning,
fatigue, etc.) In what sense then can we say that $\left(A^{1},A^{2}\right)\looparrowleft\left(\alpha^{1},\alpha^{2}\right)$,
and by what means can we find out if this is true? 

Many empirical situations have precisely the same formal structure.
In QM, an example is provided by the Bohmian version of the EPR paradigm
\cite{BohmAha1957}: two subatomic particles are emitted from a common
source in such a way that they retain highly correlated spins as they
run away from each other. An experiment may consist, e.g., in measuring
the spin of electron 1 along one of two axes, $\alpha_{1}^{1}$ or
$\alpha_{2}^{1}$, and (in another location but simultaneously in
some inertial frame of reference) measuring the spin of electron 2
along one of two axes, $\alpha_{1}^{2}$ or $\alpha_{2}^{2}$. The
outcome of a measurement on electron 1, $A^{1}$, is a random variable
with two possible values, ``up'' or ``down,'' and the same holds
for $A^{2}$, outcome of a measurement on electron 2. The question
here is: for $i=1,2$, can we say that $A^{i}$ may only depend on
$\alpha^{i}$, even though $A^{1}$ and $A^{2}$ are not independent?
What makes this situation formally identical with the double-detection
example is that the measurements along different axes, $\alpha_{1}^{i}$
and $\alpha_{2}^{i}$, are \emph{noncommuting}, i.e., they cannot
be performed on the $i$th particle simultaneously. This makes it
possible to consider them (measurements performed, not to be confused
with their recorded outcomes) as mutually exclusive values of input
$\alpha^{i}$. The results of such an experiment are described by
(\ref{eq:2-2-2}), with Yes/No interpreted as spin up/down. In the
original EPR paradigm \cite{EinPodRosen1935} the non-commuting measurements
are those of momentum and location, each with a continuum of possible
values. Our parallel with the issue of selective influences requires
that the measurements of the momentum and of the location of a given
particle be interpreted as mutually exclusive values of one and the
same input, ``(measurement of the) momentum-location of the particle.''
This may be less intuitive than the analogous interpretation of the
spins along different axes. 

The question of selective influences cannot generally be decided based
on the marginal distributions of the outputs alone. The most important
example here is the classical CHSH experiment \cite{ClauHorShiHolt1969}
where the marginal distributions of $A^{1}$ and $A^{2}$ (in the
case of two electrons) remain constant, with $\Pr\left[spin\: up\right]=\nicefrac{1}{2}$.
Examples from psychology are also readily available, especially if
one adopts a copula view of the joint distributions. Thus, $\alpha^{1}$
and $\alpha^{2}$ may represent two stimuli presented in a succession
(each having several values), and $A^{1},A^{2}$ be \emph{response
times quantiles}. The marginal distributions then are always the same,
unit-uniform.

\section{A Historical Note}

The issue of selective influences was introduced to psychology in
Sternberg's influential paper \cite{Stern1969}, in the context of
studying consecutive ``stages'' of information processing. Sternberg
acknowledged that selective influences can hold even if the durations
of the stages are not stochastically independent, but he lacked mathematical
apparatus for dealing with this possibility. Townsend \cite{Town1984}
proposed to formalize the notion of selectively influenced and stochastically
interdependent random variables by the concept of ``\emph{indirect
nonselectiveness}'': the conditional distribution of the variable
$A$$^{1}$ given any value $a^{2}$ of the variable $A^{2}$, depends
on $\alpha^{1}$ only, and, by symmetry, the conditional distribution
of $A^{2}$ at any $A^{1}=a^{1}$ depends on $\alpha^{2}$ only. Under
the name of ``\emph{conditionally selective influence}'' this notion
was mathematically characterized and generalized in \cite{Dzh1999}.
Thus, if all combinations of values of inputs $\alpha^{1},\alpha^{2}$
are allowable and random outputs $A^{1},A^{2}$ are discrete, the
diagram $\left(A^{1},A^{2}\right)\overset{cond}{\leftarrow}\left(a^{1},a^{2}\right)$,
where $\stackrel{cond}{\leftarrow}$ means ``is conditionally selectively
influenced,'' holds if and only if $\Pr\left[A^{1}=a^{1},A^{2}=a^{2}\left|\;\alpha_{x}^{1},\alpha_{y}^{2}\right.\right]$
can be presented as 
\begin{equation}
f_{12}\left(a^{1},a^{2}\right)f_{1}\left(a^{1},\alpha_{x}^{1}\right)f_{2}\left(a^{2},\alpha_{y}^{2}\right)f\left(\alpha_{x}^{1},\alpha_{y}^{2}\right),
\end{equation}
for all values $\left(a^{1},a^{2}\right)$ of $\left(A^{1},A^{2}\right)$
at all treatments $\left(\alpha_{x}^{1},\alpha_{y}^{2}\right)$. Conditional
selectivity is a useful notion, but it is not a satisfactory formalization
of the intuitive notion of selective influences. The reason is that
$\left(A^{1},A^{2}\right)\overset{cond}{\leftarrow}\left(a^{1},a^{2}\right)$
can be shown \cite{Dzh1999} to violate the following obvious property
of an acceptable definition: the marginal distributions of $A^{1}$
and $A^{2}$ do not depend on, respectively, $\alpha^{2}$ and $\alpha^{1}$
(``\emph{marginal selectivity}'' \cite{TownSchw1989}). 

A different approach to selective influences, reviewed below, is based
on \cite{Dzh2001,Dzh2003c,DzhGluh2006,DzhKuj2010,DzhKuj2012,DzhKuj_inpress,KujDzh2008}.
As it turns out%
\footnote{This was first pointed out to us by Jerome Busemeyer (personal communication,
November 2010), for which we remain deeply grateful.%
} this approach parallels the development in QM of the issue of whether
an EPR experiment can have a ``classical'' explanation (in terms
of non-contextual local variables). The Joint Distribution Criterion
which is at the heart of this development (see below) was indirectly
introduced in the celebrated work of Bell \cite{Bell1964}, and explicitly
in \cite{Fine1982a,Fine1982b,SuppesZanotti1981}.

\section{\label{sec:Theory-of-Selective}Basic Notions}

Aimed at providing a broad overview of concepts and results, the content
of this paper partially overlaps with that of several previous publications,
especially \cite{DzhKuj2012,DzhKuj_inpress,KujDzh2008}.

Random variables are understood in the broadest sense, as measurable
functions $X:V_{s}\rightarrow V$, with no restrictions on the sample
spaces $\left(V_{s},\Sigma_{s},\mu_{s}\right)$ and the induced probability
spaces (\emph{distributions}) $\left(V,\Sigma,\mu\right)$. In particular,
any set $X$ of jointly distributed random variables (functions on
the same sample space) is a random variable, and its distribution
$\left(V,\Sigma,\mu\right)$ is referred to as the \emph{joint distribution}
of its elements. We use symbol $\sim$ in the meaning of ``has the
same distribution as.'' A random variable in the narrow sense is
a special case of a random entity, with $V$ a finite product of countable
sets and intervals of reals, and $\Sigma$ the smallest sigma-algebra
containing the corresponding product of power sets and Lebesgue sigma-algebras.
Note that a vector of random variables in the narrow sense is a random
variable in the narrow sense.

Consider an indexed set $\alpha=\left\{ \alpha^{\lambda}:\lambda\in\Lambda\right\} $,
with each $\alpha^{\lambda}$ being a set referred to as a (deterministic)
\emph{input}, with the elements of $\left\{ \lambda\right\} \times\alpha^{\lambda}$
called \emph{input points}. Input points therefore are pairs of the
form $x=\left(\lambda,w\right)$, with $w\in\alpha^{\lambda}$, and
should not be confused with \emph{input values} $w$. A nonempty set
$\Phi\subset\prod_{\lambda\in\Lambda}\alpha^{\lambda}$ is called
a set of \emph{(allowable) treatments. }A treatment therefore is a
function $\phi:\Lambda\rightarrow\bigcup_{\lambda\in\Lambda}\alpha^{\lambda}$
such that $\phi\left(\lambda\right)\in\alpha^{\lambda}$ for any $\lambda\in\Lambda$.

Let there be a collection of sets of random variables $A_{\phi}^{\lambda}$
($\lambda\in\Lambda$, $\phi\in\Phi$), referred to as (random) \emph{outputs},
with distributions $\left(V^{\lambda},\Sigma^{\lambda},\mu_{\phi}^{\lambda}\right)$.
Let 
\begin{equation}
A_{\phi}=\left\{ A_{\phi}^{\lambda}:\lambda\in\Lambda\right\} ,\;\phi\in\Phi,
\end{equation}
 be a random variable with a known distribution (the joint distribution
of all $A_{\phi}^{\lambda}$ in $A_{\phi}$) for every treatment $\phi\in\Phi$.
We define 
\begin{equation}
A^{\lambda}=\left\{ A_{\phi}^{\lambda}:\phi\in\Phi\right\} ,\;\lambda\in\Lambda,
\end{equation}
 with the understanding that $A^{\lambda}$ is not generally a random
variable, i.e., $A_{\phi}^{\lambda}$ for different $\phi$ are not
necessarily jointly distributed. The definition of the relation 
\begin{equation}
\left\{ A^{\lambda}:\lambda\in\Lambda\right\} \looparrowleft\left\{ \alpha^{\lambda}:\lambda\in\Lambda\right\} ,\label{eq:SIgeneral}
\end{equation}
interpreted as ``for each $\lambda\in\Lambda$, $A^{\lambda}$ may
be influenced by $\alpha^{\lambda}$ only,'' can be given in three
equivalent forms: 
\begin{description}
\item [{($\SI_{1}$)}] there are independent random variables $C$, $\left\{ S^{\lambda}:\lambda\in\Lambda\right\} $,
and functions 
\begin{equation}
\left\{ R^{\lambda}\left(w,C,S^{\lambda}\right):w\in\alpha^{\lambda},\lambda\in\Lambda\right\} ,
\end{equation}
such that, for any treatment $\phi\in\Phi$, 
\begin{equation}
\left\{ R^{\lambda}\left(\phi\left(\lambda\right),C,S^{\lambda}\right):\lambda\in\Lambda\right\} \sim A_{\phi};
\end{equation}

\item [{($\SI_{2}$)}] there is a random variable $C$ and functions 
\begin{equation}
\left\{ P^{\lambda}\left(w,C\right):w\in\alpha^{\lambda},\lambda\in\Lambda\right\} ,
\end{equation}
such that, for any treatment $\phi\in\Phi$, 
\begin{equation}
\left\{ P^{\lambda}\left(\phi\left(\lambda\right),C\right):\lambda\in\Lambda\right\} \sim A_{\phi};
\end{equation}

\item [{($\JDC$)}] there is a set of jointly distributed random variables
\begin{equation}
H=\left\{ H_{w}^{\lambda}:w\in\alpha^{\lambda},\lambda\in\Lambda\right\} 
\end{equation}
(one random variable for every value of every input), such that, for
any treatment $\phi\in\Phi$, 
\begin{equation}
\left\{ H_{\phi\left(\lambda\right)}^{\lambda}:\lambda\in\Lambda\right\} \sim A_{\phi}.\label{eq:JDC condition}
\end{equation}

\end{description}
The latter statement constitutes the \emph{Joint Distribution Criterion}
($\JDC$) for selective influences, and $H$ is called the $\JDC$
\emph{(indexed) set}. The proof of the equivalence \cite{DzhKuj2010}
obtains essentially by the definition of a joint distribution, which
seems to have been overlooked in the earlier derivations \cite{Fine1982a,Fine1982b}.
If $\Lambda=\left\{ 1,\ldots,n\right\} $ and all outputs $A^{\lambda}$
are random variables in the narrow sense, then $C$ in $\SI_{2}$
and $C,S^{1},\ldots,S^{n}$ in $\SI_{1}$ can also be chosen to be
random variables in the narrow sense; moreover, their distribution
functions can be chosen arbitrarily, provided they are continuous
and strictly increasing on their domains, e.g., unit uniform \cite{DzhKuj2012}.

Two important consequences of (\ref{eq:SIgeneral}) are as follows:
\begin{enumerate}
\item \emph{(nestedness}) any subset $\Lambda'$ of $\Lambda$, $\left\{ A^{\lambda}:\lambda\in\Lambda'\right\} \looparrowleft\left\{ \alpha^{\lambda}:\lambda\in\Lambda'\right\} $;
in particular, $\left\{ A^{\lambda}:\lambda\in\Lambda'\right\} $
may not depend on inputs outside $\Lambda'$ (\emph{complete marginal
selectivity});
\item (\emph{invariance} \emph{with respect to input-value-specific transformations})
for any set of measurable functions $\left\{ F_{w}^{\lambda}\left(a\right):w\in\alpha^{\lambda},\lambda\in\Lambda,a\in V^{\lambda}\right\} $,
\begin{equation}
\left(B^{\lambda}:\lambda\in\Lambda\right)\looparrowleft\left\{ \alpha^{\lambda}:\lambda\in\Lambda\right\} \label{eq:trans}
\end{equation}
where $B^{\lambda}=\left\{ B_{\phi}^{\lambda}:\phi\in\Phi\right\} $,
and $B_{\phi}^{\lambda}=F_{\phi\left(\lambda\right)}^{\lambda}\left(A_{\phi}^{\lambda}\right)$.
\end{enumerate}
These properties should be viewed as desiderata for any reasonable
definition of selective influences. 

In QM, $\SI_{1}$ corresponds to the existence of a \emph{``classical''
probabilistic} explanation. In psychology, statement $\SI_{1}$ combined
with auxiliary assumptions was used in \cite{Dzh2003a} and \cite{KujDzh2009}
to analyze the representability of same-different pairwise discrimination
probabilities by means of \emph{Thurstonian-type models} in which
two stimuli being compared are mapped into random entities (distributed
in some hypothetical space of mental images) that in turn are mapped
(deterministically or probabilistically) into a response, ``same''
or ``different.'' Statement $\SI_{1}$ was also used to analyze
the response time distributions for \emph{parallel-serial networks
of mental operations} with selectively influenced components \cite{DzhSchwSung2004}.
Note that the representation of the outputs $A^{\lambda}$ as functions
of the corresponding inputs $\alpha^{i}$ and unobservable sources
of randomness, A$^{\lambda}\textnormal{-}$specific ($S^{\lambda}$)
and common ($C$), includes as special cases all conceivable generalizations
and combinations of \emph{regression} and \emph{factor analyses},
with our term ``input'' corresponding to the traditional ``regressor,''
and the term ``source of randomness'' to the factor-analytic ``factor.''
This observation alone shows the potentially unlimited sphere of applicability
of $\SI_{1}$. 

Statement $\SI_{2}$ (corresponding in QM to \emph{``classical''
determinsitic} explanation) and $\JDC$ turn out to be more convenient
in dealing with certain foundational probabilistic issues \cite{DzhGluh2006}
and for the construction of the working \emph{tests} (\emph{necessary
conditions}) for selective influences \cite{DzhKuj2010,DzhKuj2012,DzhKuj_inpress,KujDzh2008}.
The tests are discussed below.

The following is a table of correspondences between the general terminology
used in dealing with the issue of selective influences, and that of
QM in dealing with EPR. 

\begin{center}
\begin{tabular}{|c|c|}
\hline 
{\small Selective Probabilistic Causality (general)} & {\small Quantum Entanglement Problem}\tabularnewline
\hline 
\hline 
{\footnotesize observed random output} & {\footnotesize outcome of a given measurement}\tabularnewline
 & {\footnotesize on a given particle}\tabularnewline
\hline 
{\footnotesize input (factor)} & {\footnotesize set of noncommuting measurements}\tabularnewline
 & {\footnotesize on a given particle}\tabularnewline
\hline 
{\footnotesize input value} & {\footnotesize one of noncommuting measurements}\tabularnewline
 & {\footnotesize on a given particle}\tabularnewline
\hline 
{\footnotesize joint distribution criterion} & {\footnotesize joint distribution criterion}\tabularnewline
\hline 
{\footnotesize diagram of selective influences} & {\footnotesize ``classical'' explanation}\tabularnewline
\hline 
{\footnotesize representation in the form SI$_{1}$} & {\footnotesize probabilistic ``classical'' explanation}\tabularnewline
\hline 
{\footnotesize representation in the form SI$_{2}$} & {\footnotesize deterministic ``classical'' explanation}\tabularnewline
\hline 
\end{tabular}
\par\end{center}

\section{Tests for Selective Influences}

Let $H=\left\{ H_{w}^{\lambda}:w\in\alpha^{\lambda},\lambda\in\Lambda\right\} $
be a \emph{hypothetical} JDC-set, i.e., a set satisfying (\ref{eq:JDC condition})
but not necessarily jointly distributed. Denoting 
\begin{equation}
\left\{ H_{\phi\left(\lambda\right)}^{\lambda}:\lambda\in\Lambda\right\} =H_{\phi},\;\phi\in\Phi,
\end{equation}
let $\mathcal{H}$ be a \emph{set of constraints} imposed on possible
distributions of $H_{\phi}$. For instance, $\mathcal{H}$ may be
the requirement that all $H_{\phi}^{\lambda}$ be composed of Bernoulli
variables, or multivariate-normally distributed. 

A statement $S\left(H_{\phi_{1}},\ldots,H_{\phi_{s}}\right)$, with
$\phi_{1},\ldots,\phi_{s}\in\Phi$, is called a \emph{test} for the
relation (\ref{eq:SIgeneral}) under constraints $\mathcal{H}$, if 
\begin{enumerate}
\item (\emph{observability}) its truth value only depends on the distributions
of $H_{\phi_{1}},\ldots,H_{\phi_{s}}$; 
\item (\emph{non-emptiness}) it is not true for all possible distributions
of $H_{\phi_{1}},\ldots,H_{\phi_{s}}$ satisfying $\mathcal{H}$,
\item (\emph{necessity}) it is true if $H$ is jointly distributed.
\end{enumerate}
If $S\left(H_{\phi_{1}},\ldots,H_{\phi_{s}}\right)$ is false for
all distributions of $H_{\phi_{1}},\ldots,H_{\phi_{s}}$ satisfying
$\mathcal{H}$ unless $H$ is jointly distributed, the test is called
a \emph{criterion} for (\ref{eq:SIgeneral}). In the following we
assume that $\mathcal{H}$ always includes the requirement of complete
marginal selectivity: for any $\Lambda'\subset\Lambda$, the joint
distribution of $\left\{ A_{\phi(\Lambda')\cup\phi(\Lambda-\Lambda')}^{\lambda}:\lambda\in\Lambda'\right\} $
does not depend on $\phi(\Lambda-\Lambda')$. If this condition is
violated, (\ref{eq:SIgeneral}) is ruled out trivially.

\subsection{Pseudo-quasi-distance tests}

A function $d:H\times H\rightarrow\mathbb{R}$ is a \emph{pseudo-quasi-metric}
(\emph{p.q.-metric}) on $H$ if, for any $H_{1},H_{2},H_{3}\in H$,

(i) $d\left(H_{1},H_{2}\right)$ only depends on the joint distribution
of $\left(H_{1},H_{2}\right)$,

(ii) $d\left(H_{1},H_{2}\right)\geq0$,

(iii) $d\left(H_{1},H_{1}\right)=0$,

(iv) $d\left(H_{1},H_{3}\right)\leq d\left(H_{1},H_{2}\right)+d\left(H_{2},H_{3}\right)$. 

The conventional \emph{pseudometrics} (also called \emph{semimetrics})
obtain by adding the property $d\left(H_{1},H_{2}\right)=d\left(H_{2},H_{1}\right)$;
the conventional \emph{quasimetrics} are obtained by adding the property
$\Pr\left[H_{1}=H_{2}\right]<1\Rightarrow d\left(H_{1},H_{2}\right)>0$.
A conventional \emph{metric} is both a pseudometric and a quasimetric.

A sequence of input points 
\begin{equation}
x_{1}=\left(\lambda_{1},w_{1}\right),\ldots,x_{l}=\left(\lambda_{l},w_{l}\right),
\end{equation}
where $w_{i}\in\alpha^{\lambda_{i}}$ for $i=1,\ldots,l\geq3$, is
called \emph{treatment-realizable} if there are treatments $\phi^{1},\ldots,\phi^{l}\in\Phi$
(not necessarily pairwise distinct), such that 
\begin{equation}
\left\{ x_{1},x_{l}\right\} \subset\phi^{1}\textnormal{ and }\left\{ x_{i-1},x_{i}\right\} \subset\phi^{i}\textnormal{ for }i=2,\ldots,l.
\end{equation}
If a JDC-set $H$ exists, then for any p.q.-metric $d$ on $H$ we
should have 
\begin{equation}
d\left(H_{w_{1}}^{\lambda_{1}},H_{w_{l}}^{\lambda_{l}}\right)=d\left(A_{\phi^{1}}^{\lambda_{1}},A_{\phi^{1}}^{\lambda_{l}}\right)
\end{equation}
 and 
\begin{equation}
d\left(H_{w_{i-1}}^{\lambda_{i-1}},H_{w_{i}}^{\lambda_{i}}\right)=d\left(A_{\phi^{i}}^{\lambda_{i-1}},A_{\phi^{i}}^{\lambda_{i}}\right),\; i=2,\ldots,l,
\end{equation}
whence 
\begin{equation}
d\left(A_{\phi^{1}}^{\lambda_{1}},A_{\phi^{1}}^{\lambda_{l}}\right)\leq\sum_{i=2}^{l}d\left(A_{\phi^{i}}^{\lambda_{i-1}},A_{\phi^{i}}^{\lambda_{i}}\right).\label{eq:distance test}
\end{equation}
This chain inequality constitutes a \emph{p.q.-metric test} for selective
influences. If this inequality is found not to hold for at least one
treatment-realizable sequence of input points, selectivity (\ref{eq:SIgeneral})
is ruled out \cite{DzhKuj_inpress}. 

It turns out that one needs to check the chain inequality only for
\emph{irreducible} treatment-realizable sequences $x_{1},\ldots,x_{l}$,
i.e., those with $x_{1}\not=x_{l}$ and with the property that the
only subsequences $\left\{ x_{i_{1}},\ldots,x_{i_{k}}\right\} $ with
$k>1$ that are subsets of treatments are pairs $\left\{ x_{1},x_{l}\right\} \textnormal{ and }\left\{ x_{i-1},x_{i}\right\} $,
for $i=2,\ldots,l$. Inequality (\ref{eq:distance test}) is satisfied
for all treatment-realizable sequences if and only if it holds for
all irreducible sequences \cite{DzhKuj_inpress}. The situation is
even simpler if $\Phi=\prod_{\lambda\in\Lambda}W^{\lambda}$ (all
logically possible treatments are allowable). Then (\ref{eq:distance test})
is satisfied for all treatment-realizable sequences if and only if
this inequality holds for all tetradic sequences of the form $x,y,s,t$,
with $x,s\in\left\{ \lambda_{1}\right\} \times\alpha^{\lambda_{1}}$,
$y,t\in\left\{ \lambda_{2}\right\} \times\alpha^{\lambda_{2}}$, $x\not=s$,
$y\not=t$, $\lambda_{1}\not=\lambda_{2}$ \cite{DzhKuj2010}.

\emph{Order-distances} constitute a special class of p.q.-metrics,
defined as follows. Let the distribution of $H_{w}^{\lambda}\in H$
be $\left(V^{\lambda},\Sigma^{\lambda},\mu_{w}^{\lambda}\right)$.
Let 
\begin{equation}
R\subset\bigcup_{\left(\lambda_{1},\lambda_{2}\right)\in\Lambda\times\Lambda}V^{\lambda_{1}}\times V^{\lambda_{2}},
\end{equation}
and let us write $a\preceq b$ for $\left(a,b\right)\in R$. Let $R$
be a total order (transitive, reflexive, and connected). We assume
that for any $\left(\lambda_{1},\lambda_{2}\right)\in\Lambda\times\Lambda$,
$\Pr\left[H_{w_{1}}^{\lambda_{1}}\preceq H_{w_{2}}^{\lambda_{2}}\right]$
is well-defined, i.e., $\left\{ \left(a,b\right):a\in V^{\lambda_{1}},b\in V^{\lambda_{2}},a\preceq b\right\} $
belongs to the product sigma-algebra over $\Sigma^{\lambda_{1}}$
and $\Sigma^{\lambda_{2}}$. Then the function 
\begin{equation}
\D\left(H_{w_{1}}^{\lambda_{1}},H_{w_{2}}^{\lambda_{2}}\right)=\Pr\left[H_{w_{1}}^{\lambda_{1}}\prec H_{w_{2}}^{\lambda_{2}}\right],
\end{equation}
where $\prec$ is the strict order induced by $\preceq$, is well-defined,
and it is a p.q.-metric on $H$, called order-distance \cite{DzhKuj_inpress}. 

As a simple example, consider the results of a\emph{ }CHSH type experiment
with two spin axes per each of two entangled $\nicefrac{1}{2}$-spin
particles. Enumerate the spin axes $1,2$ for either particle, enumerate
the two outcomes (up and down) of each measurement $1,2$ for particle
1 and $1',2'$ for particle 2, and denote
\begin{equation}
\Pr\left[H_{i}^{1}=k,H_{j}^{2}=l'\right]=\Pr\left[A_{\left(i,j\right)}^{1}=k,A_{\left(i,j\right)}^{2}=l'\right]=p_{kl|ij},
\end{equation}
where $i,j,k,l\in\left\{ 1,2\right\} $. Define the order-distance
$\D_{1}$ by putting $1\simeq1'\prec2\simeq2'$, where $\simeq$ is
equivalence induced by $\preceq$. We have then the chain inequality
\begin{equation}
\begin{array}{l}
p_{12|12}=\D_{1}(H_{1}^{1},\! H_{2\!}^{2})\\
\quad\leq\D_{1}(H_{1}^{1},\! H_{1}^{2})\!+\!\D_{1}(H_{1}^{2},\! H_{2\!}^{1})\!+\!\D_{1}(H_{2\!}^{1},\! H_{2\!}^{2})=p_{12|11}\!+\! p_{21|21}\!+\! p_{12|22}.
\end{array}\label{eq:Q1}
\end{equation}
Consider next a similar inequality for the order-distance $\D_{2}$
defined by $1\simeq2'\prec2\simeq1'$:

\begin{equation}
\begin{array}{l}
p_{11|12}=\D_{2}(H_{1}^{1},\! H_{2\!}^{2})\\
\quad\leq\D_{2}(H_{1}^{1},\! H_{1}^{2})\!+\!\D_{2}(H_{1}^{2},\! H_{2\!}^{1})\!+\!\D_{2}(H_{2\!}^{1},\! H_{2\!}^{2})=p_{11|11}\!+\! p_{22|21}\!+\! p_{11|22}.
\end{array}\label{eq:Q2}
\end{equation}
By simple algebra, denoting
\begin{equation}
\Pr\left[H_{i}^{1}=k\right]=p_{k\cdot|i\cdot},\;\Pr\left[H_{j}^{2}=l'\right]=p_{\cdot l|\cdot j},\;
\end{equation}
the conjunction of (\ref{eq:Q1}) and (\ref{eq:Q2}) can be shown
to be equivalent to
\begin{equation}
-1\leq p_{11|11}+p_{11|21}+p_{11|22}-p_{11|12}-p{}_{1\cdot|2\cdot}-p_{\cdot1|\cdot1}\leq0.\label{eq:Fine1}
\end{equation}
One derives analogously
\begin{equation}
\begin{array}{c}
-1\leq p_{11|12}+p_{11|22}+p_{11|21}-p_{11|11}-p{}_{1\cdot|2\cdot}-p{}_{\cdot1|\cdot2}\leq0,\\
-1\leq p_{11|21}+p_{11|11}+p_{11|12}-p_{11|22}-p{}_{1\cdot|1\cdot}-p_{\cdot1|\cdot1}\leq0,\\
-1\leq p_{11|22}+p_{11|12}+p_{11|11}-p_{11|21}-p{}_{1\cdot|1\cdot}-p{}_{\cdot1|\cdot2}\leq0.
\end{array}\label{eq:Fine2}
\end{equation}
The four double-inequalities (\ref{eq:Fine1})-(\ref{eq:Fine2}) can
be referred to as the \emph{Bell-CHSH-Fine inequalities} \cite{Fine1982a,Fine1982b},
necessary and sufficient conditions for the CHSH type experiment to
have a ``classical'' explanation.

\subsection{Cosphericity Tests}

Let the outputs $A_{\phi}^{\lambda}$ all be random variables in the
narrow sense. Denote, for any distinct $\lambda_{1},\lambda_{2}\in\Lambda$
and any $\phi\in\Phi$ with $\phi\left(\lambda_{1}\right)=w_{1}$
and $\phi\left(\lambda_{2}\right)=w_{2}$,
\begin{equation}
\mathrm{Cor}\left[H_{w_{1}}^{\lambda_{1}},H_{w_{2}}^{\lambda_{2}}\right]=\mathrm{Cor}\left[A_{\phi}^{\lambda_{1}},A_{\phi}^{\lambda_{2}}\right]=\rho_{w_{1}w_{2}}^{\lambda_{1}\lambda_{2}},
\end{equation}
where $\mathrm{Cor}$ designates correlation. Let $\phi_{1},\phi_{2},\phi_{3},\phi_{4}\in\Phi$
be any treatments with
\begin{equation}
\begin{array}{cc}
\phi_{1}\left(\lambda_{1}\right)=\phi_{2}\left(\lambda_{1}\right)=w_{1}; & \phi_{1}\left(\lambda_{2}\right)=\phi_{3}\left(\lambda_{2}\right)=w_{2}\\
\phi_{4}\left(\lambda_{1}\right)=\phi_{2}\left(\lambda_{1}\right)=w'_{1}; & \phi_{4}\left(\lambda_{2}\right)=\phi_{3}\left(\lambda_{2}\right)=w'_{2}.
\end{array}
\end{equation}
Then, as shown in \cite{KujDzh2008}, if the components of $H$ are
jointly distributed,
\begin{equation}
\begin{array}{l}
\left\vert \rho_{w_{1}w_{2}}^{\lambda_{1}\lambda_{2}}\rho_{w_{1}w'_{2}}^{\lambda_{1}\lambda_{2}}-\rho_{w'_{1}w_{2}}^{\lambda_{1}\lambda_{2}}\rho_{w'_{1}w'_{2}}^{\lambda_{1}\lambda_{2}}\right\vert \\
\quad\leq\sqrt{1-\left(\rho_{w_{1}w_{2}}^{\lambda_{1}\lambda_{2}}\right)^{2}}\sqrt{1-\left(\rho_{w_{1}w'_{2}}^{\lambda_{1}\lambda_{2}}\right)^{2}}+\sqrt{1-\left(\rho_{w'_{1}w_{2}}^{\lambda_{1}\lambda_{2}}\right)^{2}}\sqrt{1-\left(\rho_{w'_{1}w'_{2}}^{\lambda_{1}\lambda_{2}}\right)^{2}},
\end{array}\label{eq:Cosphericity}
\end{equation}
This is the \emph{cosphericity test} for (\ref{eq:SIgeneral}), called
so because geometrically (\ref{eq:Cosphericity}) describes the possibility
to place four points ($w_{1},w_{2},w'_{1},w'_{2}$) on a unit sphere
in 3D Euclidean space so that the angles between the corresponding
radius-vectors have cosines equal to the correlations. Note that an
outcome of this test does not allow to predict the outcome of the
same test applied to nonlinearly input-value-specifically transformed
random variables. Due to (\ref{eq:trans}), this creates a multitude
of cosphericity tests for one and the same initial set of outputs
$A_{\phi}^{\lambda}$ . 

In the all-important for behavioral sciences $2\times2$ factorial
design ($\Lambda=\left\{ 1,2\right\} $, each input is binary, and
$\Phi$ consists of all four possible treatments), the cosphericity
test is a criterion for $\left(A^{1},A^{2}\right)\looparrowleft\left(\alpha^{1},\alpha^{2}\right)$
if (perhaps following some input-value-specific transformation) the
outputs are bivariate normally distributed for all four treatments
\cite{KujDzh2008}.

\subsection{Linear Feasibility Test}

The \emph{Linear Feasibility Test} ($\LFT$) is a criterion for selective
influences in all situations involving finite sets of inputs/outputs,
$\Lambda=\left\{ 1,\ldots,n\right\} $, with the $i$th input and
$i$th output having finite sets of values, $\left\{ 1,\ldots,k_{i}\right\} $
and $\left\{ 1,\ldots,m_{i}\right\} $, respectively \cite{DzhKuj2012}.
In other situations $\LFT$ can be used as a necessary condition because
every set of possible values can be discretized. The distributions
of $H_{\phi}=\left(H_{j_{1}}^{1},\ldots,H_{j_{n}}^{n}\right)$ are
represented by probabilities 
\begin{equation}
\Pr\left[H_{j_{1}}^{1}=a_{1},\ldots,H_{j_{n}}^{n}=a_{n}\right]=\Pr\left[A_{\phi}^{1}=a_{1},\ldots,A_{\phi}^{n}=a_{n}\right],
\end{equation}
with $\phi=\left(j_{1},\ldots,j_{n}\right)\in\Phi$ and 
\begin{equation}
\left(a_{1},\ldots,a_{n}\right)\in\left\{ 1,\ldots,m_{1}\right\} \times\cdots\times\left\{ 1,\ldots,m_{n}\right\} .
\end{equation}
We consider this probability the $\left[\left(a_{1},\ldots,a_{n}\right),\left(j_{1},\ldots,j_{n}\right)\right]$th
component of the $m_{1}\cdots m_{n}t$-vector $P$ (with $t$ denoting
the number of treatments in $\Phi$). The joint distribution of $H$
in $\JDC$, if it exists, is represented by probabilities 
\begin{equation}
\Pr\left[H_{1}^{1}=h_{1}^{1}\ldots,H_{k_{1}}^{1}=h_{k_{1}}^{1},\ldots,H_{1}^{n}=h_{1}^{n},\ldots,H_{k_{n}}^{n}=h_{k_{n}}^{n}\right],
\end{equation}
with 
\begin{equation}
\left(h_{1}^{1},\ldots,h_{k_{1}}^{1},\ldots,h_{1}^{n},\ldots,h_{k_{n}}^{n}\right)\in\left\{ 1,\ldots,m_{1}\right\} ^{k_{1}}\times\ldots\times\left\{ 1,\ldots,m_{n}\right\} ^{k_{n}}.
\end{equation}
We consider this probability the $\left(h_{1}^{1},\ldots,h_{k_{1}}^{1},\ldots,h_{1}^{n},\ldots,h_{k_{n}}^{n}\right)$th
component of the $\left(m_{1}\right)^{k_{1}}\cdots\left(m_{n}\right)^{k_{n}}$-vector
$Q$. Consider now the Boolean matrix $M$ with rows corresponding
to components of $P$ and columns to components of $Q$: let $M\left(r,c\right)=1$
if and only if 
\begin{enumerate}
\item row $r$ corresponds to the $\left[\left(j_{1},\ldots,j_{n}\right),\left(a_{1},\ldots,a_{n}\right)\right]$th
component of $P$, 
\item column $c$ to the $\left(h_{1}^{1},\ldots,h_{k_{1}}^{1},\ldots,h_{1}^{n},\ldots,h_{k_{n}}^{n}\right)$th
component of $Q$, and 
\item $h_{j_{1}}^{1}=a_{1},\ldots,h_{j_{n}}^{n}=a_{n}$. 
\end{enumerate}
Clearly, the vector $Q$ exists if and only if the system 
\begin{equation}
\mbox{\ensuremath{MQ=P,\; Q\geq0}}
\end{equation}
has a solution (is \emph{feasible}). This is a linear programming
task in the standard form (with a constant objective function). Let
$\mathcal{L}\left(P\right)$ be a Boolean function equal to 1 if and
only if this system is feasible. $\mathcal{L}\left(P\right)$ is known
to be computable, its time complexity being polynomial \cite{Karmarkar1984}.

The potential of $\JDC$ to lead to $\LFT$ and provide an ultimate
criterion for the Bohmian entanglement problem has not been utilized
in quantum physics until relatively recently, when $\LFT$ was proposed
in \cite{WernerWolf2001a,WernerWolf2001b} and \cite{BasoaltoPercival2003}.
But the essence of the idea can be found in \cite{Pitowski1989}.
Given a set of numerical (experimentally estimated or theoretical)
probabilities, computing $\mathcal{L}\left(P\right)$ is always preferable
to dealing with explicit inequalities as their number becomes very
large even for moderate-size vectors $P$. The classical Bell-CHSH-Fine
inequalities (\ref{eq:Fine1})-(\ref{eq:Fine2}) for $n=2$, $k_{1}=k_{2}=2$,
$m_{1}=m_{2}=2$ (assuming that the marginal selectivity equalities
hold) number just 8, but already for $n=2$, $k_{1}=k_{2}=2$ with
$m_{1}=m_{2}=3$ (describing, e.g., an EPR experiment with two spin-$1$
particles, or two spin-$\nicefrac{1}{2}$ ones and inefficient detectors),
our computations yield 1080 inequalitiies equivalent to $\mathcal{L}\left(P\right)=1$.
For $n=3$, $k_{1}=k_{2}=k_{3}=2$ and $m_{1}=m_{2}=m_{3}=2$, corresponding
to the GHZ paradigm \cite{GreenHornZeil1989} with three spin-$\nicefrac{1}{2}$
particles, this number is 53792. Lists of such inequalities can be
derived ``mechanically'' from the format of matrix $M$ using well-known
facet enumeration algorithms (see, e.g., program lrs at http://cgm.cs.mcgill.ca/\textasciitilde{}avis/C/lrs.html).
Once such a system of inequalities $S$ is derived, one can use it
to prove necessity (or sufficiency) of any other system $S'$ by showing,
with the aid of a linear programming algorithm, that $S'$ is redundant
when added to $S$ (respectively, $S$ is redundant when added to
$S'$).

\subsubsection*{Acknowledgments.}

This research has been supported by the NSF grant SES-1155956 to Purdue
University and the Academy of Finland grant 121855 to University of
Jyväskylä.

\label{references}


\begin{thebibliography}{10}
\bibitem{BasoaltoPercival2003} Basoalto, R.M., Percival, I.C.: BellTest
and CHSH experiments with more than two settings. Journal of Physics
A: Mathematical \& General 36, 7411\textendash{}7423 (2003)

\bibitem{Bell1964}Bell, J.: On the Einstein-Podolsky-Rosen paradox.
Physics 1, 195-200 (1964)

\bibitem{BohmAha1957}Bohm, D., Aharonov, Y.: Discussion of Experimental
Proof for the Paradox of Einstein, Rosen and Podolski. \emph{Physical
Review} 108, 1070-1076 (1957)

\bibitem{ClauHorShiHolt1969}Clauser, J.F., Horne, M.A., Shimony,
A., Holt, R.A.: Proposed experiment to test local hidden-variable
theories. Physical Review Letters 23, 880-884 (1969)

\bibitem{Dzh1999}Dzhafarov, E.N.: Conditionally selective dependence
of random variables on external factors. Journal of Mathematical Psychology
43, 123-157 (1999)

\bibitem{Dzh2001}Dzhafarov, E.N.: Unconditionally selective dependence
of random variables on external factors. Journal of Mathematical Psychology
45, 421-451 (2001)

\bibitem{Dzh2003c}Dzhafarov, E.N.: Selective influence through conditional
independence. Psychometrika 68, 7-26 (2003)

\bibitem{Dzh2003a}Dzhafarov, E.N.: Thurstonian-type representations
for ``same-different'' discriminations: Probabilistic decisions
and interdependent images. Journal of Mathematical Psychology 47,
229-243 (2003) {[}see Dzhafarov, E.N.: Corrigendum to ``Thurstonian-type
representations for `same--different' discriminations: Probabilistic
decisions and interdependent images.'' Journal of Mathematical Psychology
50, 511 (2006){]}

\bibitem{DzhGluh2006}Dzhafarov, E.N., Gluhovsky, I.: Notes on selective
influence, probabilistic causality, and probabilistic dimensionality.
Journal of Mathematical Psychology 50, 390--401 (2006)

\bibitem{DzhKuj2010}Dzhafarov, E.N., Kujala, J.V.: The Joint Distribution
Criterion and the Distance Tests for Selective Probabilistic Causality.
Frontiers in Quantitative Psychology and Measurement 1:151 doi: 10.3389\slash{}fpsyg.2010.0015
(2010)

\bibitem{DzhKuj2012}Dzhafarov, E.N., Kujala, J.V.: Selectivity in
probabilistic causality: Where psychology runs into quantum physics.
Journal of Mathematical Psychology 56, 54-63 (2012)

\bibitem{DzhKuj_inpress}Dzhafarov, E.N., Kujala, J.V.: Order-distance
and other metric-like functions on jointly distributed random variables.
Proceedings of the American Mathematical Society (in press as of 2011)

\bibitem{DzhSchwSung2004}Dzhafarov, E.N., Schweickert, R., Sung,
K.: Mental architectures with selectively influenced but stochastically
interdependent components. Journal of Mathematical Psychology 48,
51-64 (2004)

\bibitem{EinPodRosen1935}Einstein, A., Podolsky, B., Rosen N.: Can
Quantum-Mechanical Description of Physical Reality be Considered Complete?
Physical Review 47, 777\textendash{}780 (1935)

\bibitem{Fine1982a}Fine, A.: Joint distributions, quantum correlations,
and commuting observables.\emph{ }Journal of Mathematical Physics
23, 1306-1310 (1982)

\bibitem{Fine1982b}Fine, A.: Hidden variables, joint probability,
and the Bell inequalities.\emph{ }Physical Review Letters 48, 291-295
(1982)

\bibitem{GreenHornZeil1989}Greenberger, D.M., Horne, M.A., Zeilinger,
A.: Going beyond Bell's theorem. In: M. Kafatos (ed.)\emph{ }Bell\textquoteright{}s
Theorem, Quantum Theory and Conceptions of the Universe, pp. 69\textendash{}72.
Kluwer, Dordrecht (1989)

\bibitem{Karmarkar1984}Karmarkar, N.: A new polynomial-time algorithm
for linear programming.\emph{ }Combinatorica 4, 373-395 (1984)

\bibitem{KujDzh2008}Kujala, J.V., Dzhafarov, E.N.: Testing for selectivity
in the dependence of random variables on external factors. Journal
of Mathematical Psychology 52, 128--144 (2008)

\bibitem{KujDzh2009}Kujala, J.V., Dzhafarov, E.N.: Regular Minimality
and Thurstonian-type modeling. Journal of Mathematical Psychology
53, 486\textendash{}501 (2009)

\bibitem{Pitowski1989}Pitowski, I.: Quantum Probability -- Quantum
Logic. Springer, Berlin (1989)

\bibitem{Stern1969}Sternberg, S.: The discovery of processing stages:
Extensions of donders' method. Acta Psychologica 30, 276\textendash{}315
(1969)

\bibitem{SuppesZanotti1981}Suppes, P., Zanotti, M.: When are probabilistic
explanations possible?\emph{ }Synthese 48, 191-199 (1981)

\bibitem{Town1984}Townsend, J. T.: Uncovering mental processes with
factorial experiments. Journal of Mathematical Psychology 28, 363\textendash{}400
(1984)

\bibitem{TownSchw1989}Townsend, J.T., Schweickert, R.: Toward the
trichotomy method of reaction times: Laying the foundation of stochastic
mental networks. Journal of Mathematical Psycholog\emph{y} 33, 309\textendash{}327
(1989)

\bibitem{WernerWolf2001a}Werner, R.F., Wolf, M.M.: All multipartite
Bell correlation inequalities for two dichotomic observables per site.
arXiv:quant-ph\slash{}0102024v1 (2001)

\bibitem{WernerWolf2001b}Werner, R.F., Wolf, M.M.:. Bell inequalities
and entanglement. arXiv:quant-ph\slash{}0107093 v2 (2001)\end{thebibliography}
\end{document}